\documentclass[11pt]{article}
\usepackage[utf8]{inputenc}
\usepackage{authblk}
\usepackage{geometry}
\usepackage[pdftex]{hyperref}
\usepackage[pdftex]{graphicx}
\usepackage[english]{babel}
\newcommand*{\ATLASLATEXPATH}{latex/}
\usepackage{\ATLASLATEXPATH atlaspackage}
\usepackage{\ATLASLATEXPATH atlasbiblatex}
\usepackage{\ATLASLATEXPATH atlasphysics}
\title{%
\textbf{
Higgs portal vector dark matter interpretation: review of Effective Field
Theory approach and ultraviolet complete models
}}

\author{Mohamed Zaazoua  $^{1}$, Loan Truong $^{2}$, Kétévi A. Assamagan  $^{3}$, Farida Fassi  $^{1}$ \\ 
\footnotesize{
\textit{ $^{1}$ Mohammed V University in Rabat, Faculty of Science} \\
\textit{ $^{2}$ University of Johannesburg, Department of Mechanical Engineering Science} \\
\textit{ $^{3}$ Brookhaven National Laboratory (BNL)}\\
}
}

\graphicspath{{figures/}}

\usepackage{mydocument-defs}

\begin{document}

\maketitle
A review of the Higgs portal-vector dark matter interpretation of the spin-independent dark-matter nucleon elastic scattering cross section is presented, where the invisible Higgs decay width measured at the LHC is used. Effective Field Theory and ultraviolet complete models are discussed. LHC interpretations show only the scalar and Majorana  dark-matter scenarios; we propose to include interpretation for vector dark matter in the EFT and UV completions theoretical framework. In addition, our studies suggest an extension of the LHC dark matter interpretations to the sub-GeV regime.



\section{Introduction}
\label{sec:intro}
The existence of a Dark Matter (DM) component of the universe is now firmly 
established, supported by astrophysical observations \cite{Clowe_2006}. While the nature of the DM 
particles and their interactions remain an open question, 
viable candidates must lie in theories beyond the Standard Model (BSM).  A 
particularly interesting class of candidates are weakly interacting massive 
particles (WIMP). They appear in many BSM theories. Due to their 
weak-scale interaction cross-section, they can accurately reproduce the observed  DM abundance in the Universe today\cite{Feng:2008ya}.

At the LHC, experiments have explored Higgs portal scenarios in which the 
125~\GeV~Higgs boson has substantial coupling with WIMP candidates 
(such as singlet scalar S, vector V, fermion $\chi$) to induce interaction 
between WIMP and nucleon; the WIMP could be invisible decay products of the Higgs boson
~\cite{Antoniadis:2004se,ArkaniHamed:1998vp,Datta:2004jg,Kanemura:2010sh,
Djouadi:2011aa,Djouadi:2012zc,SHROCK1982250,Choudhury:1993hv,Eboli:2000ze,
Davoudiasl:2004aj,Godbole:2003it,Ghosh:2012ep,Belanger:2013kya,Curtin:2013fra}.
Therefore, limits on the branching ratio (\Bhinv) from invisible Higgs decay  can be used to set upper bounds on spin-independent DM-nucleon scattering cross section \XSwn.
LHC interpretations complement direct and indirect detection results.
\cite{Petricca:2017zdp,Behnke:2016lsk,Akerib:2016vxi,Cui:2017nnn,Aprile:2018dbl,Agnes:2018ves}. 

The Effective Field Theory (EFT) approach is based on a decription of unkown DM-Standard Model (SM) interactions in a very economical way. This has attracted 
significant attention, especially because of its simplicity and flexibility 
which allows it to be used in vastly different search contexts. For the scalar 
and Majorana fermion WIMP candidates, the EFT approach \cite{Djouadi:2011aa,Djouadi:2012zc} 
can be safely used. Hence, the EFT approach \cite{Djouadi:2011aa} is used in 
LHC Run--1 papers \cite{Aad:2015pla,Calfayan:2058131}.  Unfortunately, the validity of this 
approach for the vector-DM case has been questioned and its limitations recognized by the theoretical and experimental communities 
 \cite{BAEK2014}. Recent efforts to develop more 
model-independent approaches to DM searches stimulated this study 
\cite{Arcadi:2020jqf}, where the EFT approach is shown to result from a valid 
ultraviolet (UV) model ; therefore, EFT is viable for vector-DM interpretations. The UV completion models have been 
investigated in two scenarios: along with the EFT approaches and in a separate 
model with additional fermions \cite{DiFranzo:2015nli} . 

This note is organized as follows: common notations used in the analyses are 
presented in Section \ref{sec:com-convention}. EFT approaches and UV complete 
models are described and discussed in Sections \ref{sec:eft}, \ref{sec:obj-eft}, 
\ref{sec:re-eft} and \ref{sec:full-uv}. In Section \ref{sec:proposal}, 
 we discuss the cases of vector dark matter (VDM)-nucleon interactions.
Dark matter in the sub-GeV 
mass range is presented in Section \ref{sec:wimp-mass}. The note is summarized 
in Section \ref{sec:conclusion}.


\section{Analysis}
\label{sec:analysis}

\subsection{Common convention}
Throughout the paper, the following conventions are utilized frequently:  \label{sec:com-convention}
\begin{enumerate}
 \item $H$: 125~\GeV~Higgs boson. %
 \item $v = 246~\GeV$: Higgs field's vacuum expectation value.
 \item $m_\text{N} = 0.938~\GeV$: proton-nucleon mass.
 \item $m_\text{V}$  : vector boson mass. %
 \item $\mH = 125~\GeV$: Higgs boson mass.  
 \item $\beta_V = \sqrt{1 - 4\frac{m_V^2}{m_H^2}}$
 \item $\beta_\text{VH} = \sqrt{1 - 4\frac{m_V^2}{m_H^2}} \bigg( 1 - 
4\frac{m_V^2}{m_H^2} + 12 \frac{m_V^4}{m_H^4} \bigg) $
 \item $ \mu_{VN}^2 = \frac{m_V^2 m_N^2 }{m_V^2 + m_N^2}$: vector DM 
reduced mass. 
 \item \Bhinv : Branching ratio of \hinv, upper limit at 90\% CL of 11\% is 
used  as the result from the recently published VBF+MET analysis 
\cite{ATLAS-CONF-2020-008}.
 \item $\Gamma^{inv} (H\rightarrow VV) = \Ghinv = \Bhinv \Gamma_H^{tot} 
= 
\frac{\Bhinv}{1 - \Bhinv} \Gamma_H^{SM} $
 \item $\Gamma_H^{SM} = 0.00407~\GeV$: Higgs width at \mH = 125~\GeV
 \item $\hbar c = 1.97327\mathrm{e}^{-14}~\GeV \times cm$
 \item $f_{N} = 0.308(18)$: Higgs-nucleon form factor \cite{Hoferichter:2017olk}
\end{enumerate}

\subsection{Effective Field Theory approach}
\label{sec:eft}
In LHC Run--1 papers \cite{Aad:2015pla,Calfayan:2058131} where \hinv combination was done, the 
90\%~CL upper limit on \Bhinv was converted into 90\%~CL upper limit on 
\XSwn with WIMP being either a scalar, a fermion or a vector boson by using the 
EFT approach \cite{Djouadi:2011aa}. In the scope of this note, only the VDM 
interpretation is discussed.

This approach suggests a model-independent Lagrangian for HVV coupling 
as the following (Equation 1 of Ref. \cite{Djouadi:2011aa}):
\begin{equation}
 \mathcal{L}_V = \frac{1}{2}\mV^2 \svv + \frac{1}{4} \lambda_V (\svv)^2 + 
                 \frac{\chVV}{4} H^{\dagger}H \svv. 
\label{eqn:lag-eft}
\end{equation}

The second term in Eq. \ref{eqn:lag-eft} is for self–interaction and it is ignored;  $\lambda_{V}$ is the self interaction coupling for the vector.
The Lagrangian has only two free parameters: HVV coupling \chVV and vector 
mass \mV. Using this Lagrangian, \XSvn together with Higgs invisible decay 
width \Ghinv are derived as functions of \mV and $\lambda_{\text{VH}}$ as follow (Equations 4 and 5 of Ref. \cite{Djouadi:2011aa}):

\begin{eqnarray}
  \Gamma^{inv} ( H \rightarrow VV) = \lambda^2_{HVV} \frac{v^2 \beta_{VH} 
m_H^3}{512 \pi m_V^4} \label{eqn:run1Gamma}\\
  \XSvn_{EFT} = \lambda_{HVV}^2 \frac{m_N^2 f_N^2}{16 \pi m_H^4 (m_V + 
m_N)^2 
} \label{eqn:run1Sigma}
 \end{eqnarray}
 
Extracting the coulping $\lambda_{HVV}$ from Equation \ref{eqn:run1Gamma} and  substitute into Equation \ref{eqn:run1Sigma}, one can find a direct relation between \XSvn and \Ghinv:
\begin{equation}
    \lambda^2_{HVV}=\Gamma^{inv} ( H \rightarrow VV) \frac{512 \pi 
m_V^4}{v^2 \beta_{VH} m_H^3} \label{eqn:run1hVV} \\
\end{equation}
\begin{eqnarray}
\XSvn_{EFT} = \Gamma^{inv}(H \rightarrow VV)\frac{512 \pi m_V^4}{v^2 
\beta_{VH} m_H^3}  \times   \frac{m_N^2 f_N^2}{16 \pi m_H^4 (m_V + m_N)^2 } \nonumber\\
\XSvn_{EFT} = \Gamma^{inv}(H \rightarrow VV)\frac{32  m_V^4 m_N^2 f_N^2}{v^2 
\beta_{VH} m_H^7 (m_V + m_N)^2}  \nonumber
\end{eqnarray}
\begin{equation}
    \XSvn_{EFT} = 32\mu_{VN}^2 \Ghinv \frac{\mV^2  \mN^2 \fN^2}{v^2  \beta_{VH} \mH^7} \label{eqn:eft-XS}
\end{equation}

Using Equation \ref{eqn:eft-XS} one can transform the limit on \Bhinv 
into the vector line interpretation as in the green hashed band in Figure 9 of Ref. \cite{Aad:2015pla}. That figure
shows the ATLAS Run--1 upper limit at the 90\% CL on the WIMP--nucleon 
scattering cross section in a Higgs portal model as a function of the mass of 
the dark-matter particle, for a scalar, Majorana fermion, or 
vector-boson WIMP. 
LHC interpreted VDM limit in EFT was claimed to be
model-independent and better than limits from direct 
detection in the regime of $\mV < \frac{\mH}{2}$. However, it drew 
controversial attention which will be discussed in Section \ref{sec:obj-eft}. 
%

\subsection{Objection on EFT, first UV model}
\label{sec:obj-eft}
In the EFT approach used in LHC Run--1 \cite{Aad:2015pla}, the mass 
of the VDM was entered arbitrarly, which leads to a non-renormalisable 
Lagrangian and violation of unitarity \cite{BAEK2014}. For this reason, it is safer to consider a better 
framework, i.e. a simple UV completion with a dark Higgs sector that gives mass 
to the vector DM via spontaneous electroweak symmetry breaking (EWSB). 
The simplest renormalisable Lagrangian for the Higgs portal VDM in such a UV 
model is given by Ref.\cite{BAEK2014}:
\begin{eqnarray}
\mathcal{L}_{VDM}=-\frac{1}{4}V_{\mu\nu}V^{\mu\nu}+D_{\mu}\Phi^{\dagger}D^{\mu}\Phi -\lambda_{\Phi}(\Phi^{\dagger}\Phi-\frac{\nu_{\Phi}^{2}}{2})^{2}  -\lambda_{\Phi H}(\Phi^{\dagger}\Phi-\frac{\nu_{\Phi}^{2}}{2})(H^{\dagger}H-\frac{\nu_{H}^{2}}{2}),
\end{eqnarray}
where $\Phi$ is the dark Higgs field which generates a nonzero mass for the VDM through spontaneous  $U(1)'$ 
breaking; 
$D_{\mu}\Phi=(\partial_{u}+ig_{X}Q_{\Phi}V_{\mu})\Phi$  and $g_{X}$ is the coupling constant.\\

From the Lagrangian, one can derive the invisible branching fraction of the Higgs decay \cite{BAEK2014}:
\begin{equation}
\Ghinv=\frac{g_{X}^{2}}{32\pi}\frac{m_{H}^{3}}{m_{V}^{2}}(1-4\frac{m_{V}^{2}}{m_{H}^{2}}+12\frac{m_{V}^{4}}{m_{H}^{4}})(1-4\frac{m_{V}^{2}}{m_{H}^{2}})^{1/2},
\end{equation}
And then, the spin independent cross section of dark matter particles scattering, can be expressed as follows \cite{BAEK2014}:
\begin{equation}
\XSvn=cos^{4}(\theta)m^{4}_{H}F(m_{V},{m_{i}},\nu)\times \XSvn_{EFT},
\end{equation}
\begin{equation}
  \simeq cos^{4}(\theta)(1-\frac{m_{H}^2}{m_{2}^2})\times \XSvn_{EFT},
\end{equation}
%
%
Where $\theta$ is the mixing angle and $m_{2}$ is the mass of the dark Higgs boson.
$\XSvn_{EFT}$ is the spin independent cross section for vector DM 
particles from the EFT approach used in LHC Run--1 \cite{Aad:2015pla}.
We can see that in the case of a UV completion model, the cross section has 
at least two additional parameters, the mass of the dark Higgs boson which is 
mostly singlet-like, and the mixing angle $\theta$ between the SM Higgs and the 
dark Higgs boson.\\

\begin{figure}[ht!]
\centering
\subfloat[  \label{fig:1stModel}]  
    {     
     \includegraphics[scale=0.38]{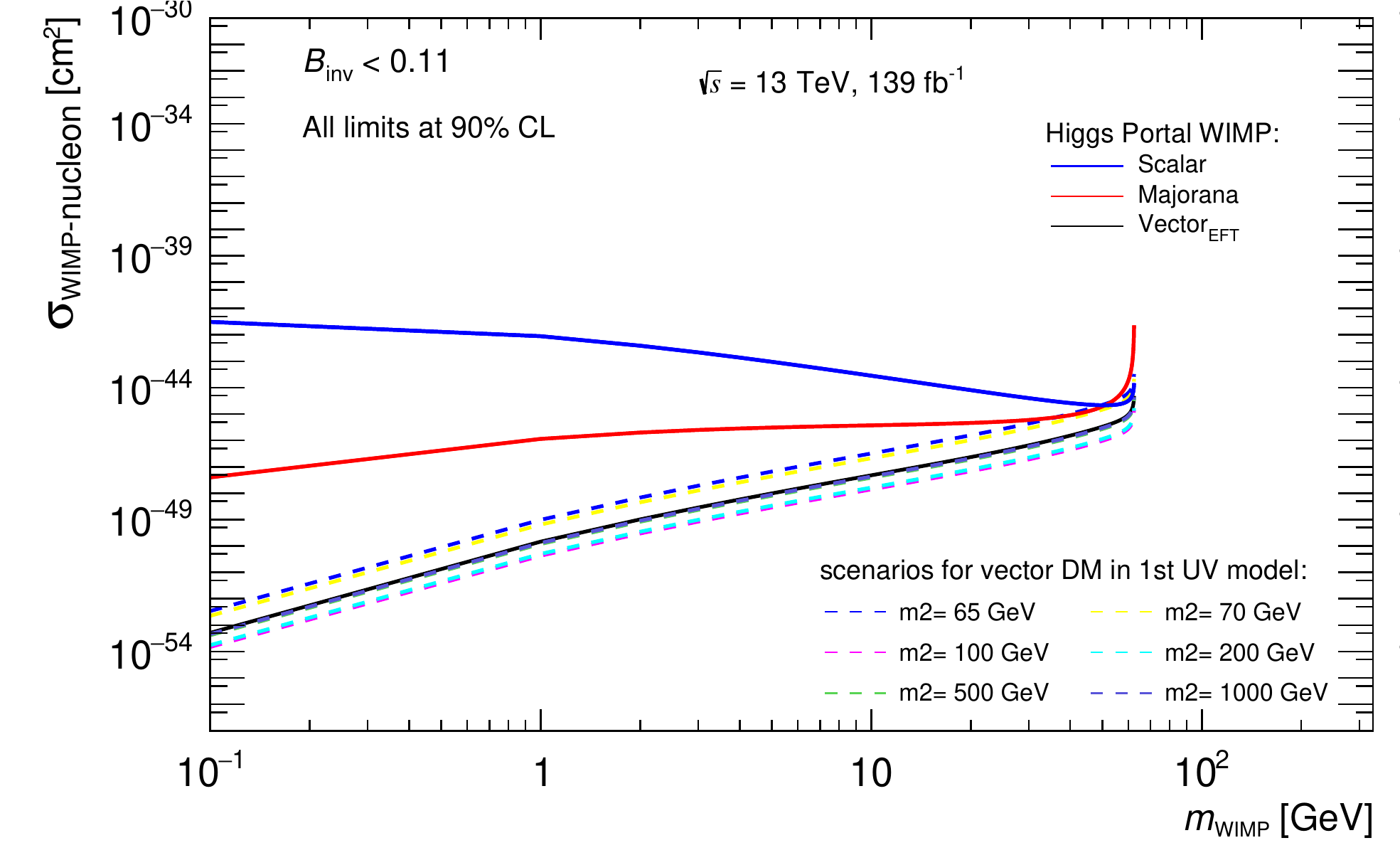}
    }
\subfloat[  \label{fig:zoming}]  
{
  \includegraphics[scale=0.39]{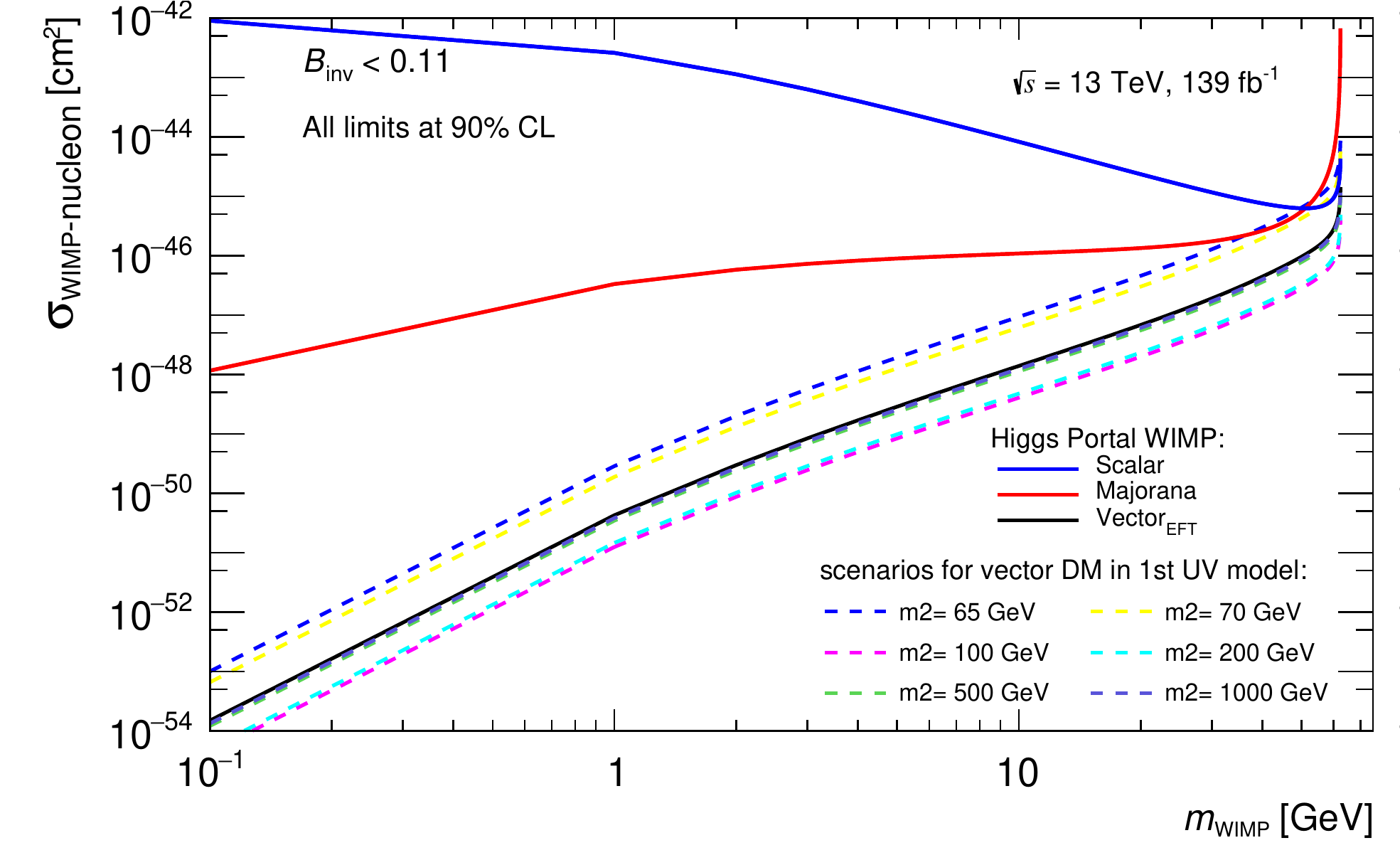}
}
\caption{ Spin independent  cross  section as function of the dark matter WIMP mass, displayed for  Scalar, Majorana and vector Higgs portal models using EFT approach.
The  vector DM state case using the first UV model is shown in Figure \ref{fig:1stModel}, for  the mixing angle $\theta=0.2$, and for the dark Higgs mass: 65, 70, 100, 200, 500, 1000 GeV  for dashed lines.
 A zoom around the vector EFT line is shown in Figure 1b to highlight the comparison between different scenarios of dark Higgs mass and the EFT approach.
}
\label{fig:1UVmodel}
\end{figure}
We investigated how the cross section evolves for the choice of small mixing and 
for different scenarios of the dark Higgs mass $m_{2}$ in the range 
$[65,1000]$ \GeV (see Figure \ref{fig:1UVmodel}).

The resulting bound on $\XSvn$ becomes weaker than the one based on EFT if the dark Higgs mass is lighter than the SM Higgs boson ($m_2=65$, 70, 100 GeV), and stronger if it is heavier than the SM Higgs boson ($m_2=200$, 500, 1000 GeV). In addtion, The UV model tends to coincide with EFT as the dark Higgs mass $m_{2}$ get larger (see Figure \ref{fig:1UVmodel}).
The usual EFT approach applies only in the case of 
$m_{2}=m_{H}cos(\theta)/\sqrt{1+cos^{2}(\theta)}$ or $m_{2} \rightarrow \infty$ 
and $\theta \rightarrow 0$. and therefore the bounds on the $\sigma^{SI}_{p}$  
should be taken with caution.


\subsection{Reanalyse EFT, second UV model}
\label{sec:re-eft}
In Ref.\cite{Arcadi:2020jqf}, theorists reanalyse the possibility that 
a Higgs-portal with a vectorial dark matter state could represent a consistent 
EFT of its UV completion. A dark Higgs sector was introduced to reproduce the 
vector mass via spontaneous electroweak symmetry breaking. Therefore the 
complete Lagrangian for dark matter phenomenology is \cite{Arcadi:2020jqf}:
\begin{eqnarray}
\mathcal{L}=\frac{1}{2}\tilde{g}M_{V}(H_{2}cos(\theta)-Hsin(\theta))V_{\mu}V^{\mu}+\frac{1}{8}\tilde{g}^{2}(H^{2}sin^{2}(\theta) -2HH_{2}sin(\theta)cos(\theta)) \nonumber\\  +H_{2}^{2}cos^{2}(\theta)V_{\mu}V^{\mu} ,
\end{eqnarray}
where $H$ is the 125 \GeV \hspace{0.05cm} SM-like Higgs boson, $H_{2}$ is the dark Higgs boson and $\tilde{g}$ the new gauge coupling.

\begin{figure}[ht!]
\centering
\subfloat[  \label{fig:2ndModel}]  
{
    \includegraphics[scale=0.4]{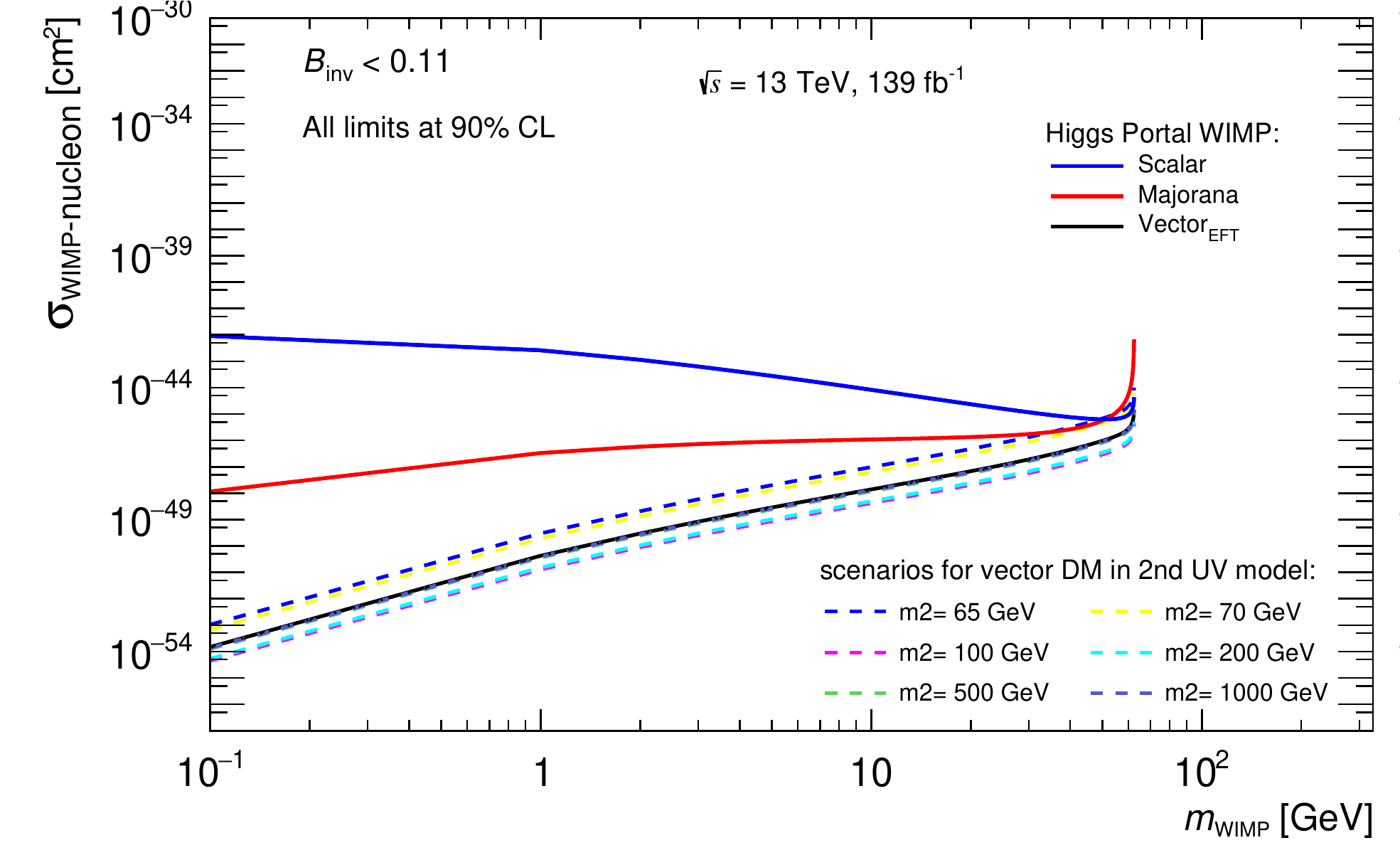}
}
\subfloat[  \label{fig:zoming2}]  
{
   \includegraphics[scale=0.4]{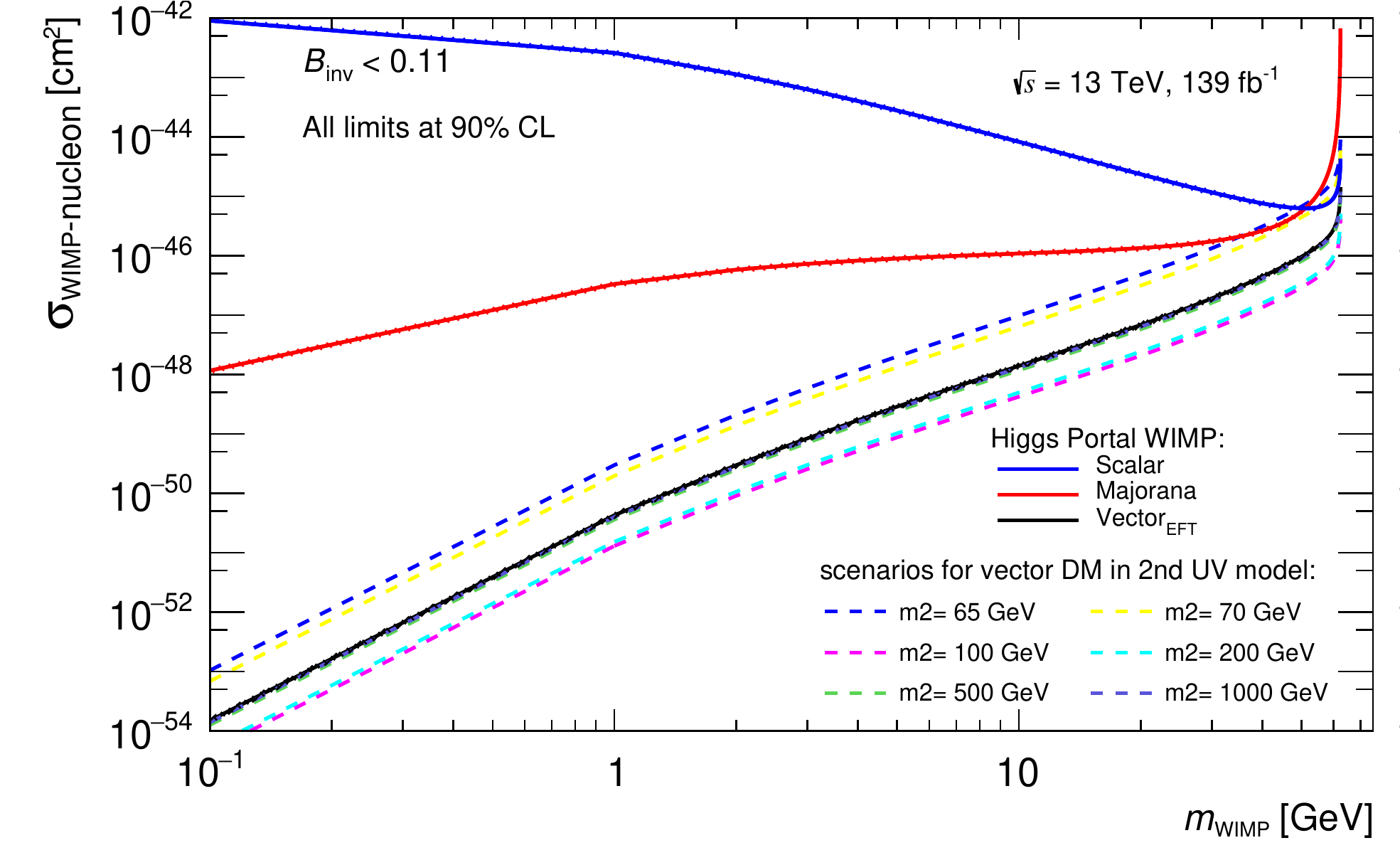}
}
\caption{Spin independent  cross  section as function of the dark matter WIMP mass, displayed for  Scalar, Majorana and vector Higgs portal models using EFT approach.
The  vector DM state case using the second UV model is shown in Figure \ref{fig:1stModel}, for  the mixing angle $\theta=0.2$, and for the dark Higgs mass: 65, 70, 100, 200, 500, 1000 GeV  for dashed lines.
 A zoom around the vector EFT line is shown in Figure 1b to highlight the comparison between different scenarios of dark Higgs mass and the EFT approach. 
}
\label{fig:2UVmodel}
\end{figure}

From the Lagrangian, one can derive the expression for $\Gamma_{inv}$ and the spin independent cross section \cite{Arcadi:2020jqf}.
\begin{equation}
(\Ghinv)_{U(1)}=\frac{\tilde{g}^{2}sin^{2}(\theta)}{32\pi}\frac{m_{H}^{3}}{m_{V}^{2}}\beta_{VH} ,
\end{equation}
\begin{eqnarray}
\XSvn=32cos^{2}(\theta)\mu_{VN}^{2}\frac{m_{V}^{2}}{m_{H}^{3}}\frac{BR(H \rightarrow VV)\Gamma^{tot}_{H}}{\beta_{VH}} \times(\frac{1}{m_{2}^{2}}-\frac{1}{m_{H}^{2}})^{2}\frac{m_{N^{2}}}{v^{2}}|f_{N}^{2}| ,
\end{eqnarray}
%
Where $\beta_{VH}$, $BR(H\rightarrow VV)\equiv \Gamma(H\rightarrow VV)/\Gamma_{H}^{tot}$, $\mu_{V_{p}}$ are the same as in Section \ref{sec:analysis} and $m_{2}$ is the dark Higgs mass.
The $\XSvn$ is different from the formula in 
Ref.\cite{Arcadi:2020jqf}. The scale was corrected from 8 to 32 after discussions with the authors of Ref.\cite{Arcadi:2020jqf}.
The prediction for VDM using EFT approach can be obtained in the limit $cos^{2}(\theta)m_{H}^{4}(1/m_{2}^{2}-1/m_{H}^{2})^{2}\approx 1$ where $sin(\theta)<<1$ and $m_{2}>>m_{H}$.\\~\\
Similarly to the  first UV model, We investigated the cross section for small mixing angles and various tuning values
of the dark Higgs boson $m_2$ in the range $[65,1000]$ \GeV (see Figure 
\ref{fig:2UVmodel}).

This exercise is extremely important not only because it shows the difference 
between the EFT and its UV completion according to values of $(\theta,m_{2})$, 
but also because it demonstrates that EFT approach could be a viable 
limit of the renormalisable model in a large region of its parameter space.

We have checked that the models introduced in Sections \ref{sec:obj-eft} and \ref{sec:re-eft} are equivalent and agree when the parameters are chosen consistently.

\subsection{Radiative Higgs portal, third UV model}
\label{sec:full-uv}
\subsubsection{Lagrangian}
\begin{figure}[ht!bp]
\centering
\includegraphics[width=0.45\textwidth]{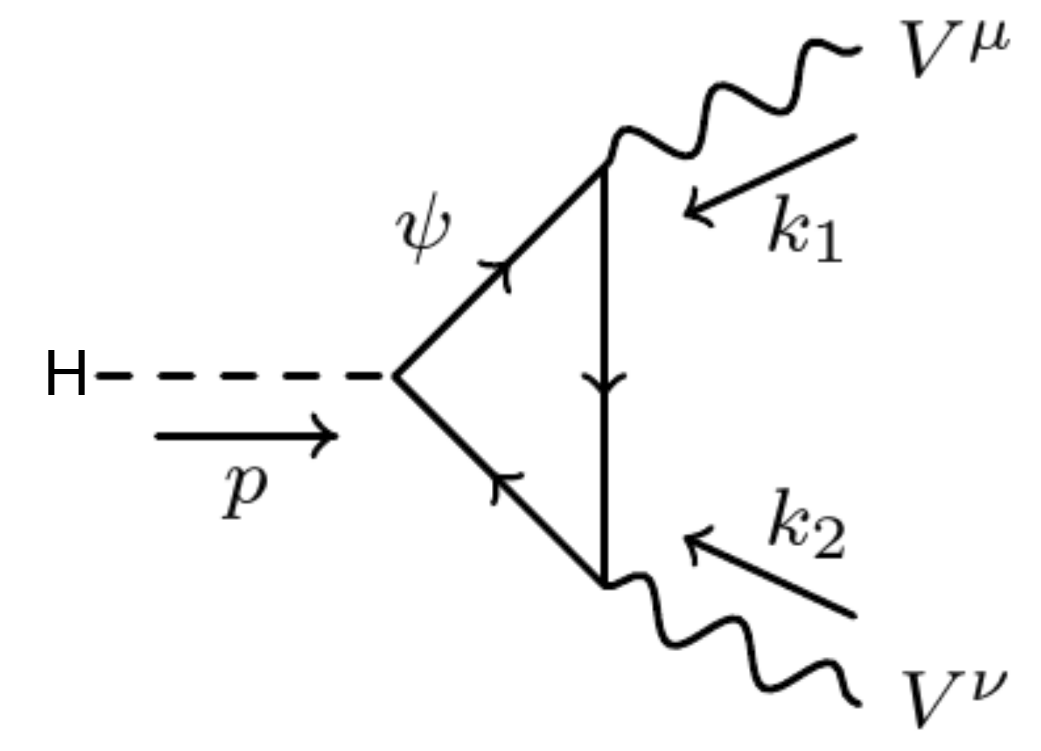}
\caption{
Fermion loop induced for HVV interaction. Figure 1 of Ref. 
\cite{DiFranzo:2015nli}
}
\label{fig:uv3-hVV}
\end{figure}

This UV model \cite{DiFranzo:2015nli} uses the same approach as 
introduced in other UV models mentioned in Sections \ref{sec:obj-eft} and 
\ref{sec:re-eft}. The vector DM is introduced as a gauge field of a 
$U(1)^{\prime}$ group which extends the SM symmetry; a Dark Higgs sector 
is added in to produce the vector boson mass via the Higgs spontaneous symmetry 
breaking mechanism. 
The Lagrangian of the vector part is as the following :
\begin{eqnarray}
 \mathcal{L} \supset -\frac{1}{4}V_{\mu\nu}V^{\mu\nu} + (D_{\mu} 
\Phi)^{\dagger} (D^{\mu} \Phi) - V(\Phi) + \lambda_P |H|^2 |\Phi|^2
\label{eqn:lag-uv3}
\end{eqnarray}
where $\lambda_P$ is the mixing parameter between the SM Higgs boson and the 
dark Higgs mode of the field $\Phi$ (Equation 2 of Ref.~\cite{DiFranzo:2015nli}). This model has a distinctive feature in 
generating the HVV coupling, and the fermions charged under SM$\times U(1)^\prime$ 
are added in, as shown below for the fermionic part of the Lagrangian :
\begin{eqnarray}
 \mathcal{L} \supset -m \epsilon^{ab} (\psi_{1a} \chi_{1b} + \psi_{2a} 
                \chi_{2b}) -m_n n_1 n_2
                -~y_{\psi} \epsilon^{ab} (\psi_{1a} H_b n_1 + \psi_{2a} H_b n_2)  \nonumber\\
                -~y_{\chi} (\chi_1 H^* n_2 + \chi_2 H^* n_1) + h.c.
 \end{eqnarray}
where $\psi, \chi, n$ are different fermion fields, a and b are $SU(2)_{W}$ indices, and H is the SM 125~\GeV ~Higgs boson (Equation 4  of Ref.~\cite{DiFranzo:2015nli}). Fermions lead to loop induced 
HVV interaction as shown in Figure \ref{fig:uv3-hVV}. 

\subsubsection{Finding relation between \XSvn and \Ghinv}
There are many different scenarios for this UV model; the studied
scenario in this note is the simplified case where the Higgs mixing parameter 
$\lambda_P \ll 1$, the charged fermions and the two heavier neutral states' 
masses are much heavier than the lightest neutral state mass, thus decouple from 
the Lagrangian. The minimal parameter space to be explored includes 
the vector mass \mV, the fermion mass \mf, the U(1)$^\prime$ coupling $g$ and 
the Yukawa coupling $y$ of the added fermion to the SM Higgs.

This model has no direct analytical relation between \Ghinv and \XSvn, and their 
computations are extensive. To obtain upper limit of \XSvn versus \mV based 
on the upper limit on \Bhinv, one has to find values of (\mf, $g$, $y$) 
which satisfy the \Bhinv upper limit within a certain precision, then calculate 
\XSvn. In our calculation, the \Bhinv limit used is 11\% at 90\%~CL from the 
recently published LHC 
analysis \cite{ATLAS-CONF-2020-008}.

Explicitly, the task requires a scan through the set (\mf, $g$, $y$) for each 
\mV point to find values of \Ghinv corresponding to \Bhinv of 11\% 
\cite{ATLAS-CONF-2020-008} within a relative precision of 0.1-1.0\%. The 
choice of 0.1-1.0\% precision is arbitrary; they are shown to have negligible 
impact on the results. Therefore, the more stringent precision of 0.1\% was considered. Some parts of the phase space can be left out of the scan since there are other constraints on those parameters:
\begin{itemize}
 \item $\mV < \frac{m_H}{2}$, as for V being on-shell decay products of the 
Higgs boson.
 \item $\mf > \frac{m_H}{2}$, to forbid the SM Higgs to decay to the additional 
fermions.
 \item $0 < g, y < 4\pi$, as rule of thumb for dimensionless couplings 
satisfying perturbation.
 \item $0 < g^2y < 40$, a model constraint \cite{DiFranzo:2015nli}.
\end{itemize}
%
All (\mf, $g$, $y$) sets that satisfy the  corresponding  11\% of the \Bhinv are used to  construct a band of \XSvn versus \mV. 
Different coarse 
to fine scanning steps of 0.1 to 0.01 on ($g$, $y$) are performed while keeping 
the same step of 1~\GeV~for \mV and 5~\GeV~for \mf, as shown in Table 
\ref{tab:vfscan}.


%
\begin{table}[ht!bp]
\begin{center}
\caption{Scanning configurations for \mV and \mf, in context of the UV model 
in Ref. \cite{DiFranzo:2015nli}
\vspace{6pt}
\label{tab:vfscan}}
\begin{tabular}{ c | c c c c }
\toprule
Variable   &First bin   &Last bin   &Step\\
\midrule
\mV (\GeV) &1          &62         &1      \\
\mf (\GeV) &64         &499        &5      \\
\bottomrule
\end{tabular}
\end{center}
\end{table}

\textbf{Coarse scan}\\
Scanning steps of 0.1 on ($g$, $y$) are performed while keeping the same step 
of 1~\GeV~for \mV and 5~\GeV~for \mf as shown in Table \ref{tab:vfscan}. Detailed 
configurations for this scan can be found in Table \ref{tab:coarse-scan}. A 
relative precision of 0.1\% on \Ghinv is required.
\begin{table}[ht!bp]
\begin{center}
\caption{Scanning configurations in the coarse scan for $g$ and $y$ in the 
context of UV model in Ref. \cite{DiFranzo:2015nli}.
\vspace{6pt}
\label{tab:coarse-scan}}
\begin{tabular}{ c | c c c c }
\toprule
Variable   &First bin   &Last bin   &Step\\
\midrule
$g$         &0          &12         &0.1\\
$y$         &0          &12         &0.1\\
\bottomrule
\end{tabular}
\end{center}
\end{table}
%
%

\textbf{Fine scan}\\
Scanning steps of 0.01 on ($g$, $y$) are performed while keeping the same step 
of 1~\GeV~for \mV and 5~\GeV~for \mf as shown in Table \ref{tab:vfscan}. 
Detailed configurations for this scan can be found in Table \ref{tab:fine-scan}. 
A relative precision of 1\% on \Ghinv is required.
\begin{table}[ht!bp]
\begin{center}
\caption{Scanning configurations in the fine scan for $g$ and $y$ in the 
context of the UV model in Ref. \cite{DiFranzo:2015nli}.
\vspace{6pt}
\label{tab:fine-scan}}
\begin{tabular}{ c | c c c c }
\toprule
Variable   &First bin   &Last bin   &Step\\
\midrule
$g$         &0          &12         &0.01\\
$y$         &0          &12         &0.01\\
\bottomrule
\end{tabular}
\end{center}
\end{table}
%
%


%
For both scans, all the found (\mf, $g$, $y$) for each \mV 
point are used to calculate \XSvn. The cross section values are then sorted 
from lowest to highest for each \mV point and plotted in Figure 
\ref{fig:uv3-scan}. 
Discussion about the plots is presented in the next section.


\subsubsection{Results}
\begin{figure}[ht!bp]
\centering
\includegraphics[width=0.49\textwidth]{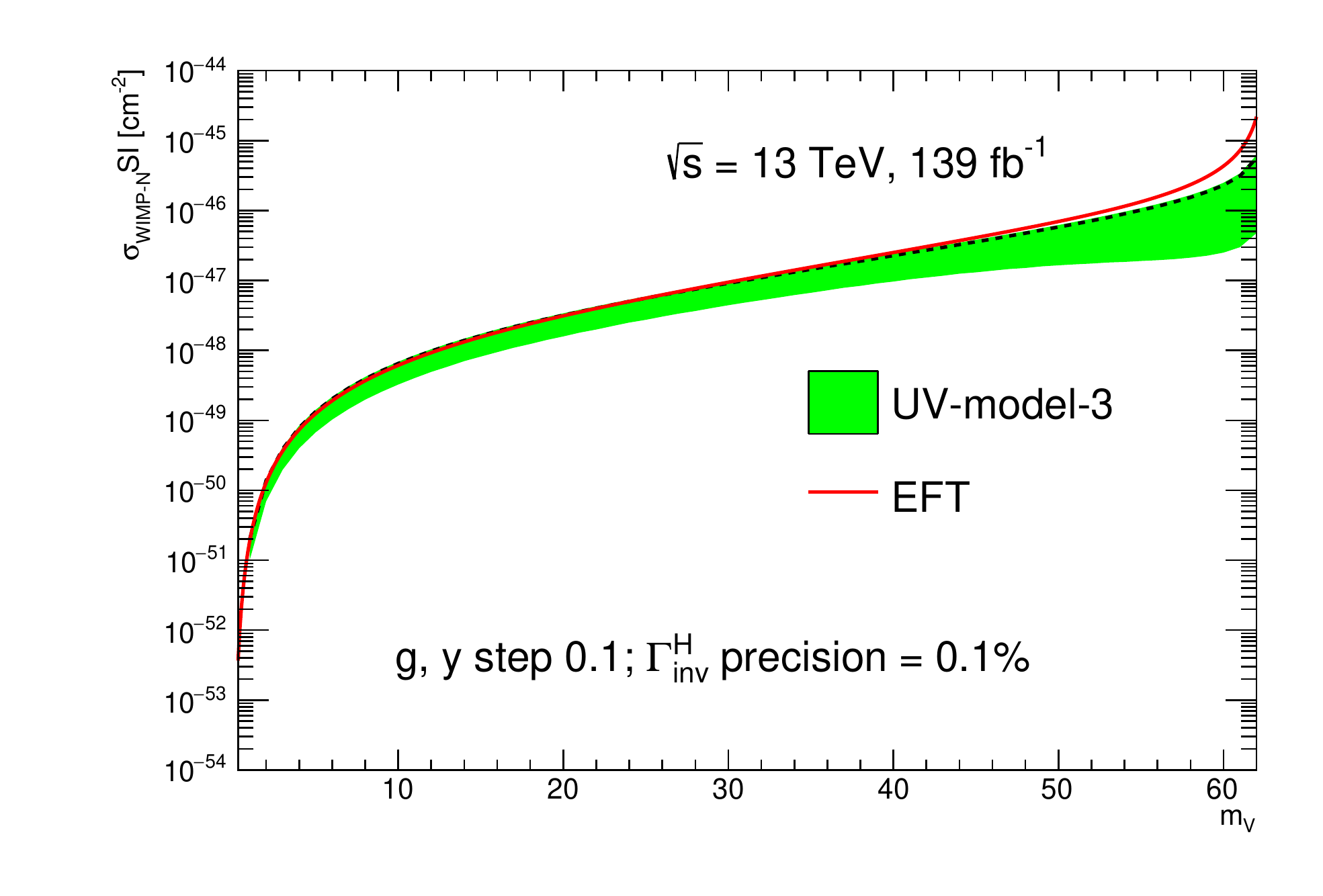}
\includegraphics[width=0.49\textwidth]{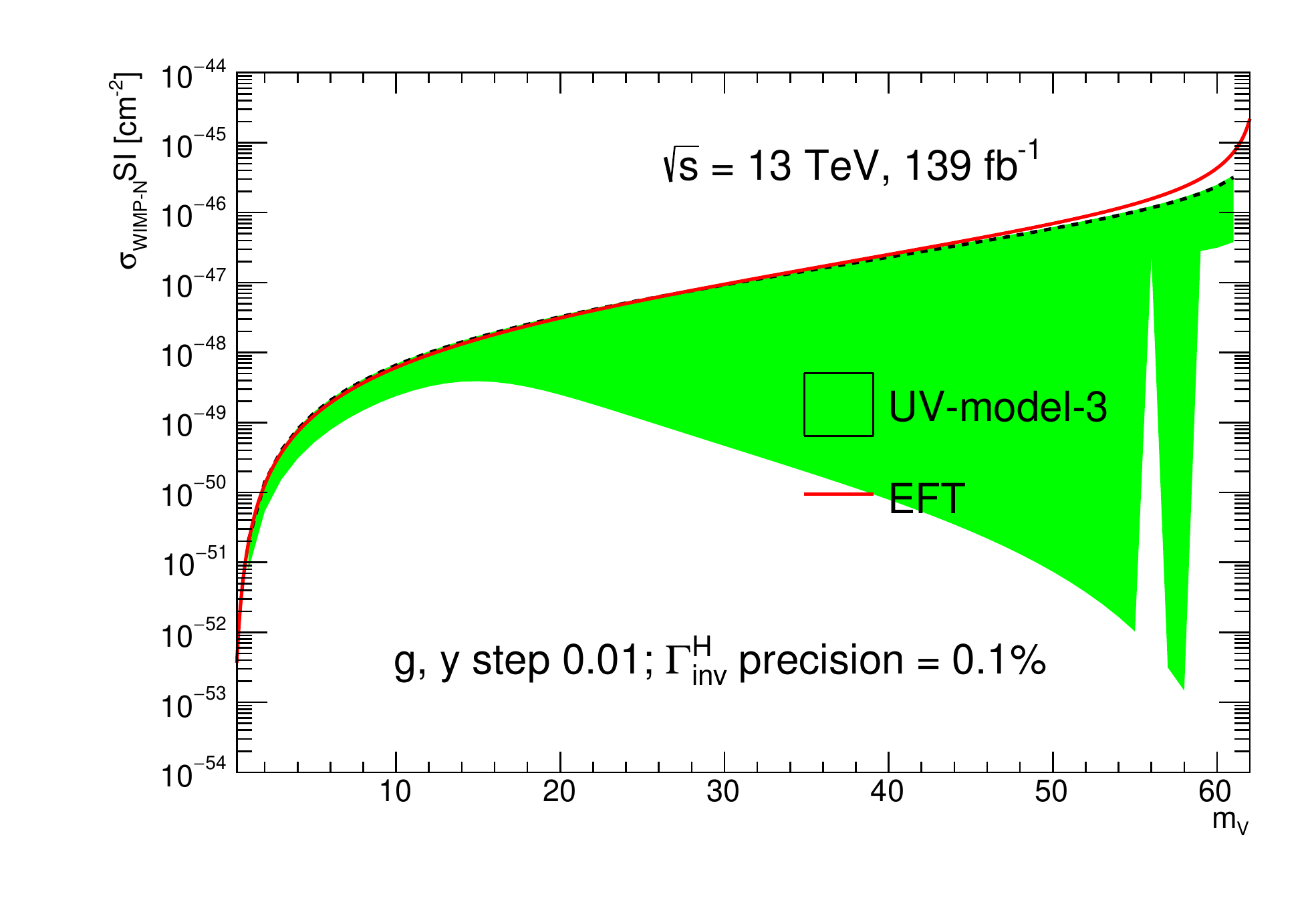}
\includegraphics[width=0.49\textwidth]{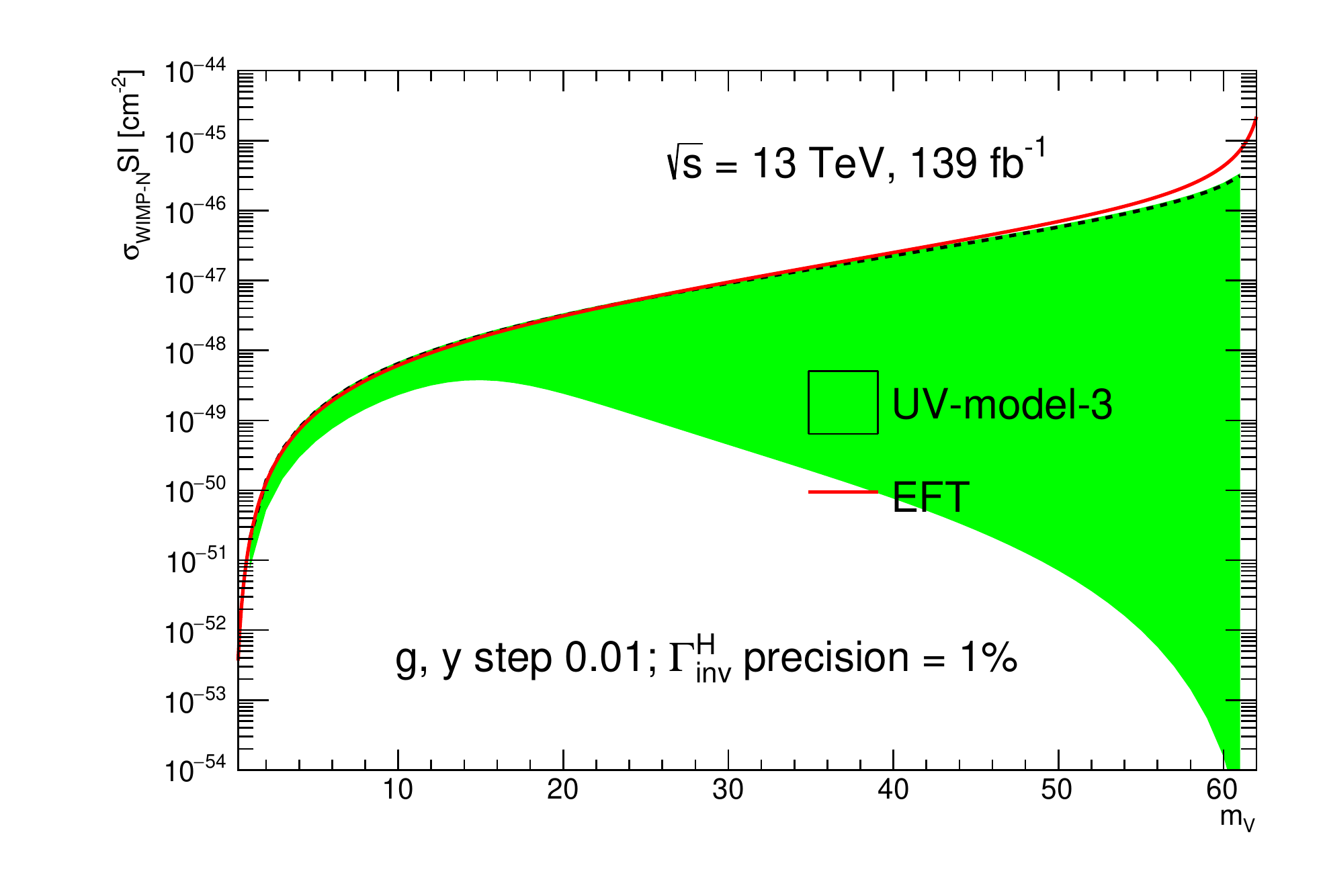}
\caption{
Green bands of upper limit on \XSvn from coarse scan in Table \ref{tab:coarse-scan} (upper left canvas), fine scan from Table \ref{tab:fine-scan}
(upper right) and fine scan  from Table \ref{tab:fine-scan} with looser precision of \Ghinv (down canvas) 
are shown in comparison with EFT red line, for the UV model in Ref. 
\cite{DiFranzo:2015nli}.
}
\label{fig:uv3-scan}
\end{figure}

\begin{figure}[ht!]
\centering
\includegraphics[width=0.49\textwidth]{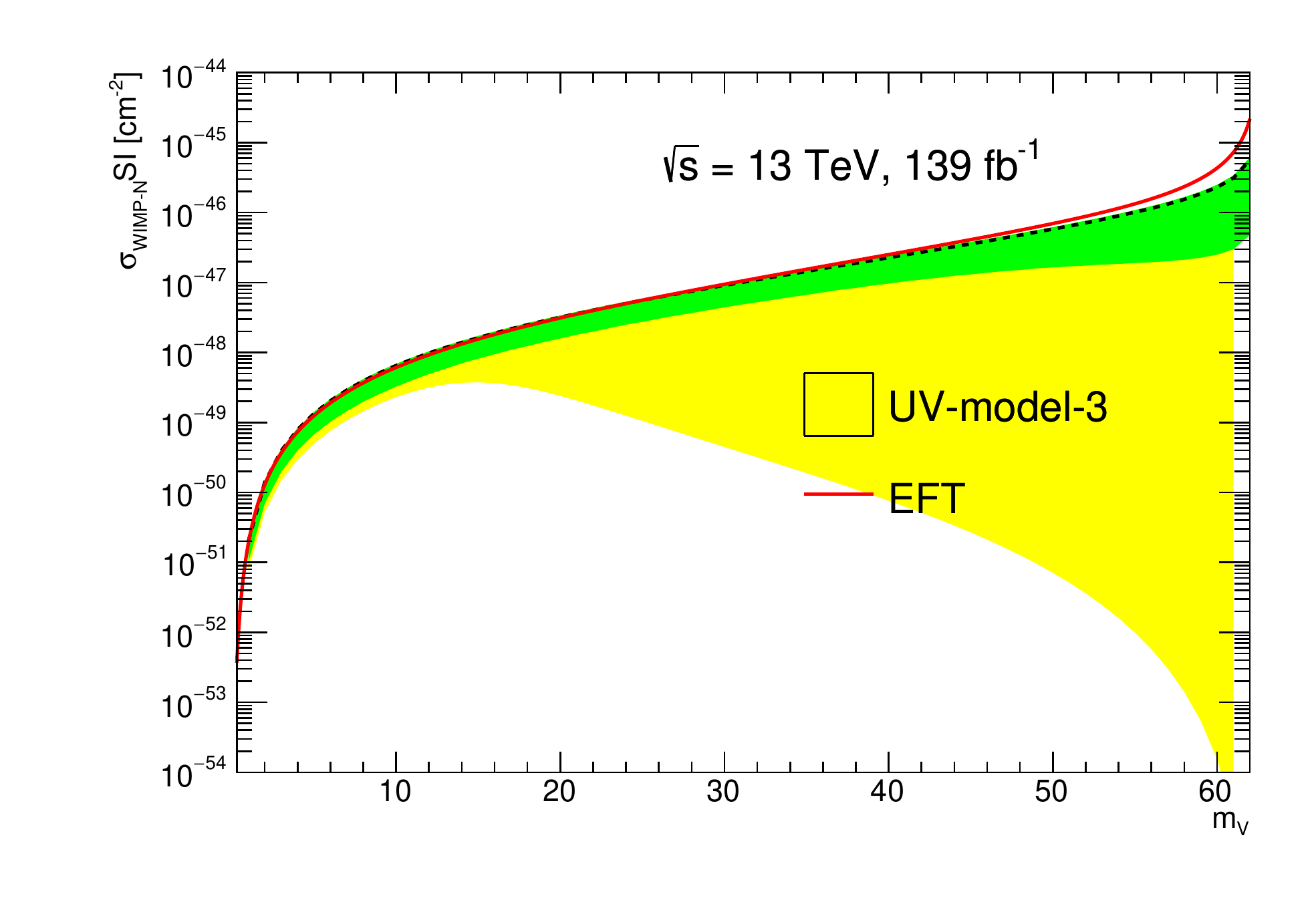}
\includegraphics[width=0.49\textwidth]{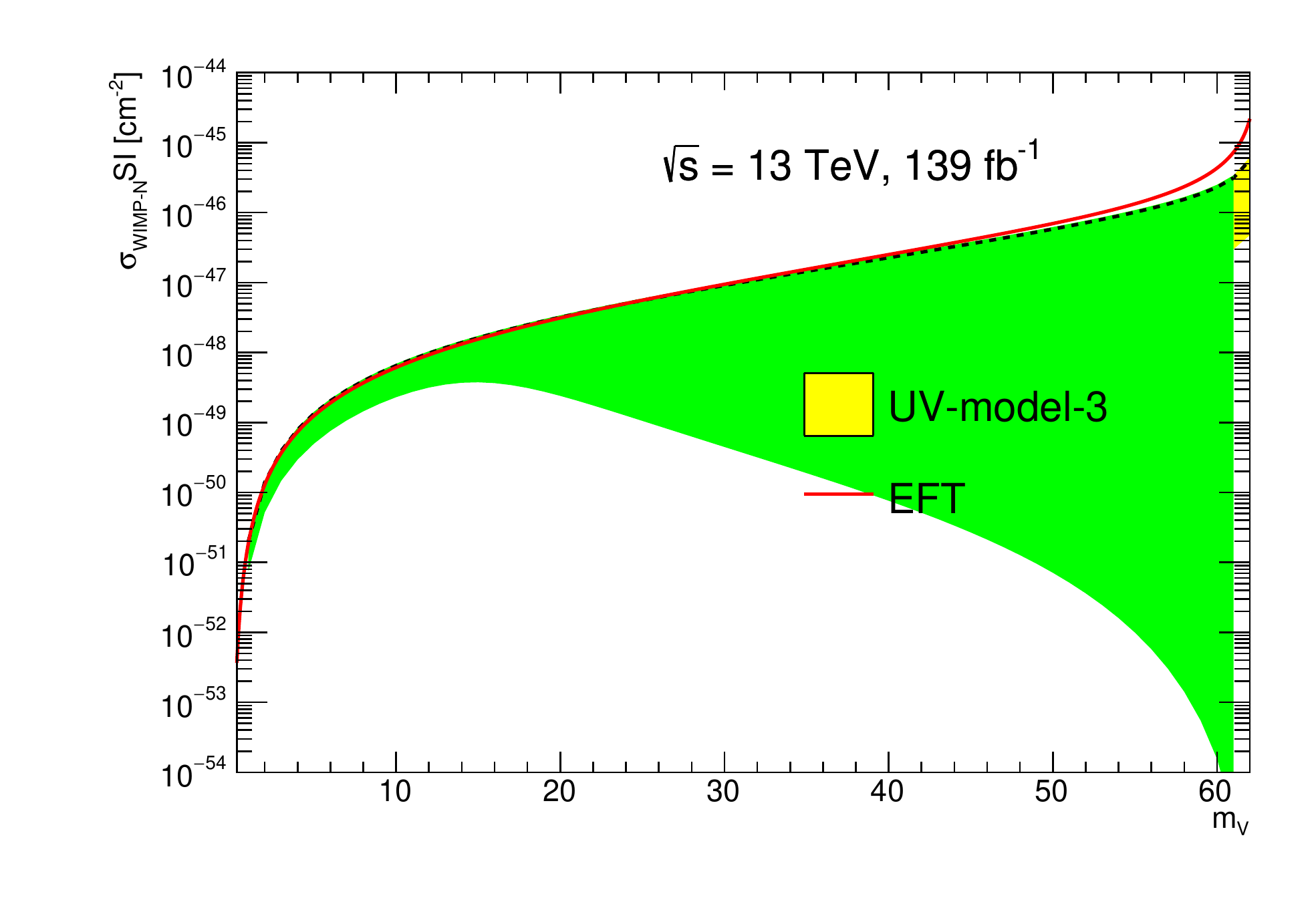}
\caption{
Superimposition of the interpretations for a coarse scan on top of a fine scan 
(left canvas) and vice versa (right canvas), for the UV model in Ref. 
\cite{DiFranzo:2015nli}.
}
\label{fig:uv3-sumperimpose}
\end{figure}

Figures \ref{fig:uv3-scan} and \ref{fig:uv3-sumperimpose} show  that 
the precision on \Ghinv does not affect the upper bound on the \XSvn  as the 
dashed lines remain the same for all cases, and stay very close to the EFT 
limit. However, as seen in the second and third plots of Figure 
\ref{fig:uv3-scan}, the fine scanning of ($g$, $y$) extends the lower bound of 
the green bands meaning that going finer in ($g$, $y$) one can achieve much 
better limits on \XSvn compared to EFT limit.

\section{Proposal}
\label{sec:proposal}
In this section we present our proposal of the Higgs portal VDM interpretation 
of the spin-independent dark-matter nucleon elastic scattering cross section 
using the invisible Higgs decay width. We propose to re-introduce the VDM limits in the LHC Higgs portal DM interpretation plots. This proposal is motivated by the results 
presented in Section \ref{sec:analysis} and could be split in three parts.

Firstly, 
The limitations of EFT approach as violation of unitarity and 
non-renormalisable Lagrangian (claimed in Section \ref{sec:obj-eft}) is 
refuted by the recent review which derived the EFT Lagrangian from a certain 
UV model as shown in Section \ref{sec:re-eft}.  This shows that the EFT approach is viable in the limit of a heavy additional scalar and small mixing angle.

Secondly, we propose to show the worst and best case scenarios of of the models described in Sections 2.3 and 2.4.

Thirdly, we propose to display the upper bound line of the UV-Model-3 discussed in Section 2.5, as shown with cyan in Figure 6.

\begin{figure}[ht!bp]
\centering
\includegraphics[width=0.9\textwidth]{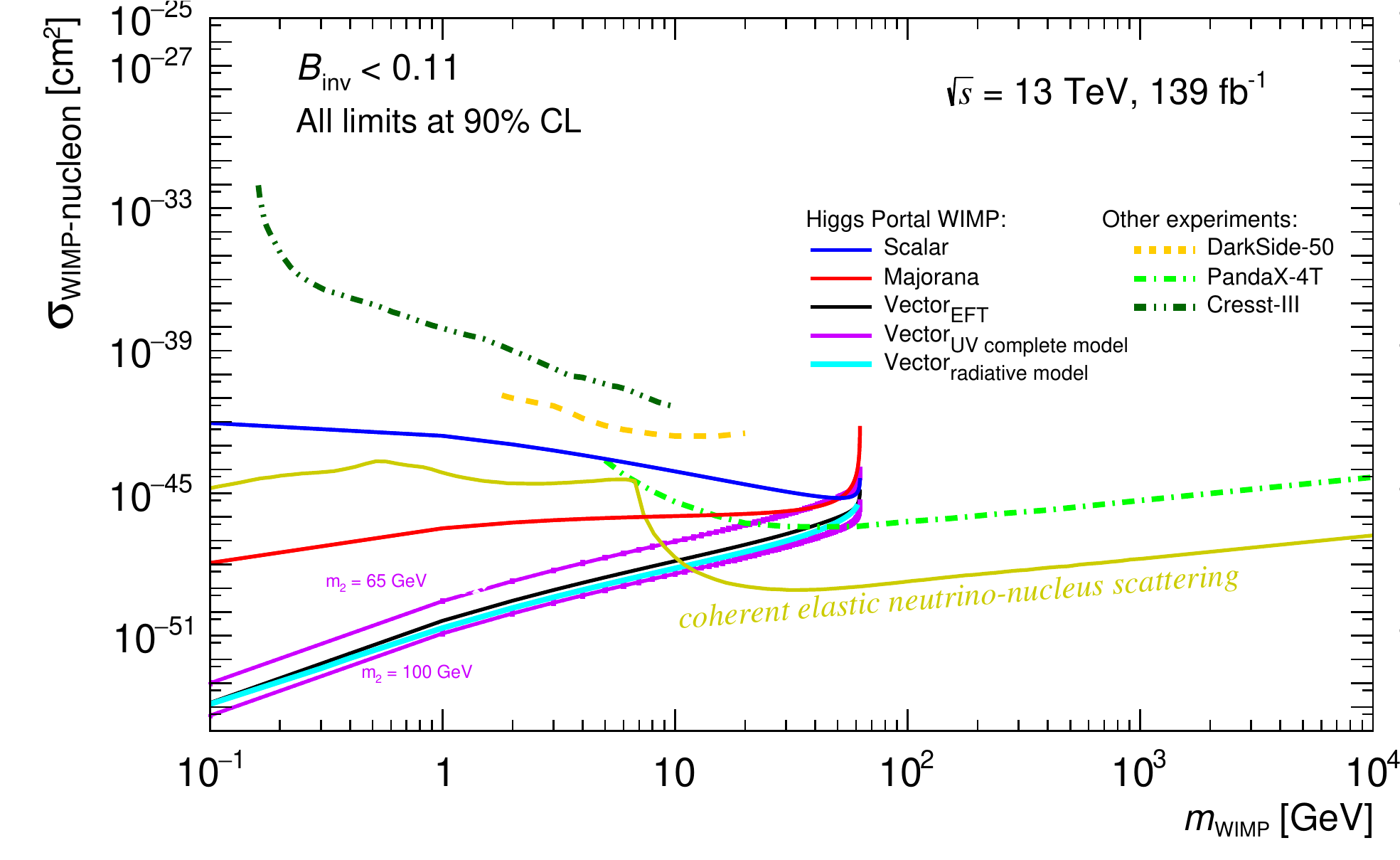}
\caption{
Upper limits on spin-independent WIMP-nucleon cross section using Higgs portal interpretations of $B_{inv}$ at 90\% CL as a function of the WIMP mass for scalar, majorana and vector states. For the vector hypothesis, the interpretation from EFT, UV complete and radiative models are presented respectively in black, magenta and cyan colours. 
Two scenarios are displayed for the UV complete model corresponding to the best and worst limits giving the mass of the dark Higgs in the range [65,1000] \GeV.
Results from direct detection experiments  and the neutrino floor for coherent elastic neutrino-nucleus scattering  are added for comparison \cite{Abdelhameed:2019hmk,darkside2018,meng2021dark,billard:tel-03259707}.    
}
\label{fig:pro-final}
\end{figure}

Our full proposal is shown in Figure \ref{fig:pro-final},
where the interpretation of the radiative Higgs portal 
(third UV model)  compared with EFT limit and with best and worst limit from the first UV model in $m_2$ range range of $[65,1000]$ GeV. Also the most stringent limits currently available from direct detection experiments are shown for comparison \cite{Abdelhameed:2019hmk,darkside2018,meng2021dark}. 
The neutrino floor for coherent elastic neutrino-nucleus scattering of astrophysical neutrino is added in \cite{neurtino-2014,neutrino-2-2014,billard:tel-03259707}. 


   \centering

\clearpage


\section{Extension to Sub-GeV WIMP masses}
\label{sec:wimp-mass}
\raggedright
The LHC Higgs-portal DM interpretation of \XSwn has been so far shown for \mV 
ranging from 1~\GeV~to $\frac{m_H}{2}$. 
The upper edge at  $\frac{m_H}{2}$ is for WIMP candidates to be produced 
on-shell from a Higgs decay. Whereas the lower edge at 1~\GeV~is
arbitrarily coming from different considerations. 

The first consideration is about the theoretical or cosmological constraint on the 
WIMP mass. However, Particle Data Group 2019 review on DM shows  the possibility 
of going to sub-\GeV~regime in many BSM models with WIMP paradigm \cite{PDG:2019}. Sections 
26.6.2 and 26.6.3 of the PDG review discussed  solid-state 
cryogenic detector experiment such as CRESST-III \cite{Abdelhameed:2019hmk}  which probes DM mass down to $\sim 160~\MeV$. 

LHC Dark Matter Working Group (LHCDMWG) white paper 
\cite{ALBERT2019100377} has recommendations for interpretation of simplified DM 
models which have s-channel spin-1 mediators decaying to fermions (invisible, 
aka DM candidates). To predict the relic density, the LHCDMWG recommends to work
under the assumption that the DM annihilation cross section of the predicted 
models is fully responsible for the DM number density \cite{ALBERT2019100377}.  
That leads to Figures 3 and 4 of Ref. \cite{ALBERT2019100377} to have DM mass
lower bound at few~\GeV. However, the mentioned bench mark models do not include
Higgs portal scenario in which the scalar Higgs boson is the mediator.

The second consideration is about the uncertainty on the \XSwn calculation via a 
Higgs mediator for LHC interpretation in the WIMP sub-\GeV~mass regime. That 
calculation depends on the coupling of the Higgs boson to a single nucleon, 
first calculated in Ref. \cite{SHIFMAN1978443} and further improved 
in Ref. \cite{Hoferichter:2017olk} whose \fN value of 0.308(18) is then used in 
Refs. \cite{ATLAS-CONF-2020-008,HIGG-2018-54}. These calculations use  lattice QCD 
formalisms which are valid continuously from negative momentum transfer to 
positive momentum transfer, thus valid for 0-momentum transfer (our case of 
WIMP-nucleon elastic scattering).
Interactions with the main author of Ref. 
\cite{Hoferichter:2017olk} resolves the consideration about \fN vs sub-\GeV  
mass. 

In conclusion, the aforementioned considerations are  not relevant to limit the LHC Higgs portal interpretations above 1~\GeV.  Therefore, we propose to show in the LHC Higgs portal interpretation plot, WIMP masses down to 0.1~\GeV---as shown in Figure 2 of Ref.~\cite{Arcadi:2020jqf}.


\section{Conclusion}
\label{sec:conclusion}
Several approaches for the interpretation of \XSvn in Higgs-portal DM 
scenarios are presented. EFT approach is reviewed and shown to be safe to be 
reinserted in the LHC Higgs portal interpretation plot. Three UV models are studied, their results 
all are shown in different parameter phase spaces. In the first two UV 
models \cite{BAEK2014,Arcadi:2020jqf}, EFT is recovered when getting 
limits in certain region of their parameter phase spaces. Whereas for the 
third UV model \cite{DiFranzo:2015nli}, result in a simplified regime is 
better than the EFT approach limits. Therefore our final proposal for the LHC Higgs portal interpretation plot is to reinsert the EFT VDM line, include the upper bound of the 
third UV model, and the worst-best lines of the first and second UV models. 
Additionally, WIMP masses in the sub-\GeV~regime are discussed and proposed to be 
extended to 0.1~\GeV~in the LHC Higgs portal interpretation plot.
\clearpage


\printbibliography

\clearpage

\end{document}